\documentclass[preprint,12pt]{elsarticle}
\usepackage{amsmath,amssymb}
\usepackage{graphicx}
\usepackage{booktabs}
\usepackage{multirow}
\usepackage{array}
\usepackage{makecell}
\usepackage{float}
\usepackage{subcaption}
\usepackage{natbib}
\usepackage{hyperref}
\usepackage{siunitx}
\usepackage{xcolor}
\usepackage{appendix}

\makeatletter
\def\ps@pprintTitle{%
	\let\@oddhead\@empty
	\let\@evenhead\@empty
	\def\@oddfoot{}%
	\let\@evenfoot\@oddfoot}
\makeatother

\begin{document}
	
	\begin{frontmatter}
		
		\title{A Variance-Based Convergence Criterion in Neural Variational Monte Carlo for Quantum Systems}
		
		\author{Huan-Chen Shi}
		\author{Er-Liang Cui\corref{cor1}}
		\ead{erliang.cui@nwafu.edu.cn}
		\author{Dan Zhou\corref{cor1}}
		\ead{danzhou@hebtu.edu.cn}
		
		\affiliation[1]{organization={College of Science, Northwest A\&F University},
			addressline={},
			city={Yangling},
			postcode={712100},
			country={China}}
		\affiliation[2]{organization={School of Physics, Hebei Normal University},
			addressline={},
			city={Shijiazhuang},
			postcode={050024},
			country={China}}
		
		\cortext[cor1]{Corresponding authors.}
		
		\begin{abstract}
			
			The optimization of neural wave functions in variational Monte Carlo crucially relies on a robust convergence criterion. While the energy variance is theoretically a definitive measure, its practical application as a primary convergence criterion has been underexplored. In this work, we develop a lightweight, general-purpose solver that utilizes the energy variance as a convergence criterion. We apply it to several systems—including the harmonic oscillator, hydrogen atom, and charmonium hadron—for validating the variance as a reliable diagnostic, and using a empirical threshold \(10^{-3}\) as the energy variance convergence values for performing rapid parameter scans to enable preliminary physical verification. To clarify the scope of our approach, we derive an inequality that delineates the limitations of variance-based optimization in nodal systems. Despite these limitations, the energy variance proves to be a highly valuable tool, guiding our solver to efficient and reliable results across a range of quantum problems.
			
		\end{abstract}
		
		
	\end{frontmatter}
	
		
		
		\section{Introduction}
		\label{sec1}
		
		The accurate description of quantum many-body systems remains a central challenge in modern physics, owing to the exponential growth of the Hilbert space with system size and the nonlinear complexity of interaction terms. Artificial Neural Networks (ANNs), established as universal function approximators by the universal approximation theorem \cite{cybenko_approximation_1989}, have opened new avenues for addressing such high-dimensional problems. Their integration with deep learning and quantum Monte Carlo (QMC) methods \cite{foulkesQuantumMonteCarlo2001,2010JPCM22b3201N,austin2012quantum} has demonstrated remarkable potential for advancing the study of quantum systems.
		
		Early pioneering studies showed that ANNs could successfully solve differential equations and eigenvalue problems, including applications to quantum systems such as the Morse potential and muonic atoms \cite{lagarisArtificialNeuralNetwork1997, lagarisArtificialNeuralNetworks1998}. The field expanded significantly with the introduction of neural-network-based variational Monte Carlo (VMC), which leveraged deep learning to approximate wave functions across a broad range of quantum systems. A landmark contribution by Carleo and Troyer introduced neural network wave functions (NNWFs) for spin models, enabling accurate calculations of ground states and dynamics in one- and two-dimensional systems \cite{carleoSolvingQuantumManybody2017a}. Parallel efforts extended these ideas to bosonic systems, quantum many-body physics, and high-dimensional partial differential equations \cite{saito2017solving, han2017deep, hanSolvingHighdimensionalPartial2018a,deng2017quantum,gao2017efficient,hanSolvingManyelectronSchrodinger2019,szabo2020neural}. Subsequent developments incorporated physical symmetries and targeted excited states \cite{chooSymmetriesManyBodyExcitations2018, ye2019neural}, while further applications demonstrated the versatility of NNWFs in describing non-equilibrium dynamics, frustrated magnets, and topological phases \cite{czischek2018quenches, glasser2018neural, kaubruegger2018chiral, saito2018method, fabiani2019investigating, schmitt2020quantum}.
		
		A major breakthrough came with the development of architectures tailored for fermionic statistics, such as FermiNet and PauliNet \cite{pfauInitioSolutionManyelectron2020, hermannDeepneuralnetworkSolutionElectronic2020}. By explicitly embedding physical constraints like the Pauli exclusion principle, these models achieved near-exact results for molecules and solids, in some cases surpassing advanced coupled-cluster methods in strongly correlated regimes. Their scalability further underscores their promise for high-precision electronic structure calculations \cite{spencerBetterFasterFermionic2020, nordhagen2023efficient}. At the same time, the emergence of robust software frameworks such as NetKet and DeepQMC has lowered technical barriers and accelerated the adoption of these methods by the broader community \cite{vicentiniNetKet3Machine2022, schatzleDeepQMCOpensourceSoftware2023}.
		
		Recent innovations continue to broaden the scope of NNWFs. Notable advances include self-attention-based architectures (e.g., Psiformer) for large molecular systems, generalized VMC algorithms for excited states and spectroscopic properties, and symmetry-adapted ansatzes for lattice models \cite{glehnSelfAttentionAnsatzAbinitio2023, pfauAccurateComputationQuantum2024, vieijra2021many,renGroundStateMolecules2023,li2024computational}. In parallel, research into optimization challenges—such as rugged energy landscapes and sign-structure difficulties—has inspired more robust training strategies and network designs \cite{bukov2021learning, park2022expressive, viteritti2022accuracy,pu_solving_2023}. Applications have also extended to real-time dynamics, supermagnonic propagation, and non-Born-Oppenheimer physics \cite{fabiani2021supermagnonic, gutierrez2022real, hofmann2022role, zhang2025schrodingernet}, further underscoring the versatility of the neural-network paradigm. Collectively, these efforts establish neural networks as a powerful and scalable framework for tackling quantum many-body problems across physics and chemistry.
		
		Despite these advances, a persistent challenge remains in verifying the convergence of ANNs-based quantum simulations. In most existing studies, convergence is judged heuristically, for example by monitoring the total energy or the behavior of parameter gradients. While such practices are convenient, they do not always provide a reliable indication of proximity to an eigenstate, particularly in systems with flat optimization landscapes or when reference energies are unavailable. This motivates the search for a more direct and physically grounded convergence measure.
		
		In this work, we investigate the energy variance as such a criterion. Since the variance of the local energy vanishes exactly for any eigenstate, it provides a rigorous and universal standard. Our theoretical analysis shows that in non-fermionic ground states of Hermitian Hamiltonians—where the wave function has no nodes and the local energy distribution exhibits no heavy tails—the variance serves as both a physically meaningful optimization objective and a quantitative convergence measure. For nodal states, variance minimization may become unstable; nevertheless, the variance remains a valuable diagnostic of convergence. We therefore employ the variance as a convergence indicator, confirming the reliability of the obtained excited-state solutions.
		
		To demonstrate its practical utility, we integrate this criterion into a lightweight neural solver and apply it to a variety of benchmark systems. Across examples including the harmonic oscillator, hydrogen atom, charmonium hadron, we show that achieving a low variance correlates with high-fidelity solutions. Furthermore, by embedding this criterion into automated parameter scans, we provide a practical tool for the preliminary verification of quantum Hamiltonians.
		
		The work is organized as follows: Sec. \ref{Sec2} proposes a numerically stable estimate of the node manifold and introduces the energy–variance criterion. Sec. \ref{Sec3} validates its reliability across multiple systems and demonstrates its utility in automated parameter scans. Sec. \ref{Sec4} concludes with a summary and outlook for future research.

		\section{Theory} \label{Sec2}
		
		The variational principle provides the theoretical foundation for optimizing trial wave functions in quantum Monte Carlo (QMC) methods. It states that the ground-state energy corresponds to the global minimum of the energy functional
		
		\begin{equation}
			E[\psi] = \frac{\langle \psi | \hat{H} | \psi \rangle}{\langle \psi | \psi \rangle},
		\end{equation}
		where \(\hat{H}\) is the Hamiltonian operator. By minimizing this functional with respect to variational parameters, one can approach the ground-state eigenfunction. A related property of the Hamiltonian is the zero-variance principle: for an exact eigenstate, the variance of the local energy vanishes,
		
		\begin{equation}
			\sigma_E^2 = \langle \psi | \hat{H}^2 | \psi \rangle - \big(\langle \psi | \hat{H} | \psi \rangle\big)^2 = 0.
		\end{equation}
		
		This suggests that variance minimization could, in principle, serve as an alternative optimization criterion. However, in practice, the statistical convergence of variance estimates is far more delicate than that of energy expectations, particularly in stochastic sampling schemes.
		
		\subsection{Statistical estimators in variational Monte Carlo}
		
		In variational Monte Carlo (VMC), the local energy is defined as
		
		\begin{equation}
			E_L(\mathbf{R}) = \frac{\hat{H}\psi(\mathbf{R})}{\psi(\mathbf{R})},
		\end{equation}
		where \(\mathbf{R}\) denotes the many-body configuration. For a set of \(N\) statistically independent samples, the sample mean and variance of the local energy are
		
		\begin{equation}
			\bar{E} = \frac{1}{N}\sum_{i=1}^N E_L^{(i)}, \quad
			\sigma^2 = \frac{1}{N-1}\sum_{i=1}^N \big(E_L^{(i)} - \bar{E}\big)^2.
		\end{equation}
		
		The standard error of the mean is governed by the second central moment,
		
		\begin{equation}
			SE(\bar{E}) = \frac{\sigma}{\sqrt{N}}, \quad \mu_2 = \sigma^2 = \langle (E_L - \bar{E})^2 \rangle,
		\end{equation}
		whereas the standard error of the variance depends on the fourth central moment\cite{mood1974},
		
		\begin{equation}
			SE(\sigma^2) \approx \sqrt{\frac{\mu_4 - \sigma^4}{N}}, \quad \mu_4 = \langle (E_L - \bar{E})^4 \rangle.
		\end{equation}
		
		Thus, the stability of energy estimates requires only that \(\mu_2\) be finite, while variance estimates demand finiteness of \(\mu_4\). In heavy-tailed distributions, the fourth moment is more prone to divergence, leading to unstable variance optimization.
		
		\subsection{Singular behavior near nodal manifolds}
		
		For systems with nodal manifolds, where \(\psi = 0\), the local energy can exhibit singular behavior in the vicinity of the nodes\cite{foulkesQuantumMonteCarlo2001,ceperley1991fermion,mitas2006structure}. Such singularities often give rise to heavy-tailed statistical distributions, which in turn complicate variance-based convergence.  
		
		Consider a nodal manifold of codimension \(k\) embedded in a \(d\)-dimensional configuration space. Let \(\delta\) denote the normal distance to the nodal manifold. In the neighborhood of the node, the wave function and the local energy scale as follows:  
		
		\begin{equation}
			\psi \sim \delta^\beta, \quad E_L \sim \delta^{-\gamma},
		\end{equation}
		where \(\beta\) characterizes the order at which the wave function vanishes, and \(\gamma\) quantifies the strength of the singularity. For typical fermionic nodes, one finds \(\beta = 1\) and \(\gamma = 1\).  
		
		Under importance sampling with weight \(|\psi|^2\), the probability density near the node scales as \(\delta^{2\beta}\). Taking into account the volume element of the nodal neighborhood, which scales as \(\delta^{k-1} d\delta\), the contribution to the \(p\)-th central moment of the local energy is given by  
		
		\begin{equation}
			\mu_p \sim \langle E_L^p \rangle \sim \int_0^{\varepsilon} \delta^{\,2\beta + k - 1 - p\gamma}\, d\delta.
		\end{equation}
		
		This integral converges as \(\varepsilon \to 0\) if and only if  
		
		\begin{equation}
			2\beta + k - 1 - p\gamma > -1,
		\end{equation}
		
		which can be rearranged into the condition  
		
		\begin{equation}
			2\beta > p\gamma - k. \label{NumericalStabilityEstimationOfNodalManifolds}
		\end{equation} 
		
		For the common case \((\beta,\gamma)=(1,1)\) and codimension \(k=1\), the condition reads \(2 > p - 1\). Thus:
		
		For \(p=2\) (standard error of the mean energy), the condition is satisfied, ensuring convergence.
		
		For \(p=4\) (standard error of the variance), the condition fails, implying divergence of the fourth moment and instability of variance estimates.
		
		If uniform sampling is used instead of \(|\psi|^2\) importance sampling, the probability density near the node is constant, and the convergence condition reduces to
		
		\begin{equation}
			0 > p\gamma - k.
		\end{equation}
		
		In this case, even the second-order moment may diverge for \(k=1\), which indicates that the energy estimate is more unstable in the absence of importance sampling suppression.
		
		Eq.~(\ref{NumericalStabilityEstimationOfNodalManifolds}) left-hand side represents the suppression of nodal singularities by the wave function, while the right-hand side quantifies the severity of the divergence. Thus, improving stability amounts need to enhancing the nodal suppression or reducing the effective singularity. Higher codimension nodes are less problematic: the effective weight of the nodal region decreases with increasing \(k\), improving the convergence of higher-order moments. This aligns with physical intuition: the more “sparse” the nodal set, the weaker its statistical impact on sampled observables.

		\subsection{Limitations and practical implications}
		
		Building on the above analysis, it is natural to distinguish between nodal and non-nodal states. In non-nodal states (\(\gamma=0\)), corresponding to non-fermionic ground states of Hermitian Hamiltonians, the local energy distribution exhibits no heavy tails and all central moments remain finite. In this regime, the variance of the local energy is not only mathematically well-defined but also physically meaningful: excited states necessarily introduce nodes and thus larger variances, so minimizing the variance naturally drives the optimization toward the ground state. Thus, in practice, the energy variance is directly employed as the loss function for non-nodal ground states.  
		
		In nodal states ($\gamma>0$), by contrast, the divergence of higher-order moments—particularly the fourth moment required for stable variance estimation—can pose significant challenges. When nodal manifolds occupy a substantial portion of the configuration space (e.g., in fermionic systems with extensive nodal surfaces), frequent statistical outliers emerge, manifesting as pronounced noise in variance estimates. Nevertheless, the energy variance remains a valuable convergence indicator: its overall descending trend reliably signals systematic improvement of the trial wave function. This is particularly useful in the later stages of optimization, once the wave function has approximately captured the correct nodal structure, as the variance provides a sensitive measure of convergence that often reveals improvements not apparent from energy monitoring alone. It is worth emphasizing that if technical measures are employed to render the nodal singularities integrable—thereby satisfying the inequality in Eq.~(\ref{NumericalStabilityEstimationOfNodalManifolds})—the energy variance can indeed be successfully utilized as a stable optimization objective. Accordingly, for nodal excited states we adopt the Schrödinger-equation residual as the primary loss function \cite{lagarisArtificialNeuralNetwork1997,lagarisArtificialNeuralNetworks1998,jin2022physics,mutukCornellPotentialNeural2019,mutukNeuralNetworkStudy2019,mutukNeuralNetworkStudy2020,mahmood2024solving,breviAddressingNonperturbativeRegime2024}, while using the variance trend as a convergence diagnostic.
		
		It should be emphasized that the inequality in Eq.~(\ref{NumericalStabilityEstimationOfNodalManifolds}) is derived under an idealized nodal-scaling assumption. While one could, in principle, circumvent its statistical implications through technical interventions such as non-physical sampling (e.g., using $|\psi|^\alpha$ with $\alpha \gg 2$) or constructing regularized observables, such strategies typically sacrifice computational efficiency, introduce additional singularities, or both. Furthermore, our analysis does not incorporate additional singularities such as electron–electron or electron–nucleus cusps. Nonetheless, these complexities do not invalidate our central conclusion: the energy variance, despite its limitations for robust optimization in nodal systems, remains a reliable and physically meaningful criterion for assessing convergence, as its systematic decrease consistently reflects the overall improvement of the trial wave function.

		\section{Results and Discussion} \label{Sec3}
		
		\subsection{Validation of the energy variance criterion}
		
		In this section, we present numerical experiments to evaluate the practical performance of the energy–variance convergence criterion. To examine its universality and reliability, we consider three representative benchmark systems that span different interaction types and spatial dimensions: the two-dimensional harmonic oscillator, the hydrogen atom, and the charmonium system.

		\subsubsection{Computational setup and training parameters}
		
		A unified neural network architecture—comprising six hidden layers with 128 neurons each—was employed across all calculations. Ground state calculations utilized an energy variance loss function, while excited states were treated using a loss based on the residual of the Schrödinger equation. The activation function was selected according to the physical characteristics of each system: Gaussian activation was used for ground states, the harmonic oscillator, and charmonium systems, whereas Tanh activation was applied to the excited states of the hydrogen atom. Optimization was performed using the AdamW optimizer \cite{loshchilov2017decoupled} with an initial learning rate of 0.001, combined with a LambdaLR learning rate scheduler. To ensure a consistent and rigorous evaluation across all 12 benchmark systems, each was trained with 8,000 walkers for 8,000 steps under a full-batch training strategy, without using weight initialization or regularization techniques.
		
		All computations were implemented in PyTorch and executed on a system equipped with an Intel Core i7‑13700K CPU and an NVIDIA GeForce RTX 4060 Ti GPU. Training time varied significantly with system complexity but remained highly manageable. For instance, a three‑dimensional single‑particle system with 1,000 walkers and 4,000 steps converged in about one minute, while a nine‑dimensional three‑body quantum dot system converged under comparable settings in roughly three minutes. These results indicate that, although our validation protocol was intentionally conservative, the solver itself is highly efficient and well suited for rapid exploration at relatively low computational cost. Moreover, evaluating the energy variance introduces no additional bottleneck: its computational cost scales linearly with the number of Monte Carlo samples, essentially matching that of the energy expectation value.
		
		It is also important to note that the core architecture of this computational framework is designed for generality; as such, it does not incorporate specialized structures to account for system-specific features like the cusp condition in Coulomb potentials. This design choice, however, serves to highlight the practical utility of the energy variance convergence criterion. Even under a non-optimal computational setup, the criterion reliably and robustly indicates when the best possible solution attainable within the current constraints has been reached. This robustness is paramount for the rapid, automated preliminary verification of physical models, which is a primary objective of our method. 
		
		\subsubsection{Results and analysis of multiple physical systems} \label{Validation}
		
		For the harmonic oscillator and hydrogen atom, atomic units are employed, with the Hamiltonians given by:
		\begin{equation}
			\hat{H}_{\mathrm{HO}}=-\frac{1}{2m}\nabla ^2+\frac{1}{2}m\omega^2r^2,
		\end{equation}
		
		\begin{equation}
			\hat{H}_{\mathrm{Hyd}}=-\frac{1}{2}\nabla ^2-\frac{1}{r},
		\end{equation}
		where $m=\omega^2=1.0$ in the harmonic oscillator of our setting.

		For the charmonium hadron states, natural units are adopted, and we solve the radial Schrödinger equation:
		\begin{equation}
			-\frac{1}{2\mu} \left[ \frac{d^2u(r)}{dr^2} - \frac{l(l+1)}{r^2} u(r) \right] + V_{c\bar{c}}(r) u(r) = E_r u(r).
		\end{equation}
		
		The effective potential form is chosen as follows\cite{mahmood2024solving,barnesHigherCharmonia2005,PhysRevD.90.054001}:
		\begin{equation}
			V_{c\bar{c}}(r) = C_f \frac{\alpha_s}{r} + br + \boldsymbol{s}_i \cdot \boldsymbol{s}_j \frac{32\pi \alpha_s}{9m_c^2} \left( \frac{\sigma}{\sqrt{\pi}} \right)^3 e^{-\sigma^2 r^2},
		\end{equation}
		
		\begin{equation}
			\boldsymbol{s}_i \cdot \boldsymbol{s}_j = \frac{s(s+1)}{2} - \frac{3}{4},
		\end{equation}
		here, \( u(r) = rR(r) \) is the reduced radial wave function, and \( l \) is the orbital angular momentum quantum number. For the sake of simplicity, the weak hyperfine structure is neglected here. For this potential model, the parameters are chosen as\cite{mahmood2024solving,barnesHigherCharmonia2005,PhysRevD.90.054001}: \( C_f = -4/3 \), \( \alpha_s = 0.5461 \), \( b = 0.1425\ \mathrm{GeV}^2 \), \( \sigma = 1.0946\ \mathrm{GeV} \), and \( m_c = 1.4796\ \mathrm{GeV} \).
		
		The calculated energies and variances for all the aforementioned systems are summarized in Table \ref{tab:results}.
		
		\begin{table}[htbp]
			\centering
			\caption{Validation of computational results for different quantum systems}
			\label{tab:results}
			\begin{tabular}{cccccc}
				\toprule
				\multirow{2}{*}{Systems} & \multirow{2}{*}{States} & \multicolumn{2}{c}{Energy} & \multirow{2}{*}{Relative Error} & \multirow{2}{*}{Energy Variance} \\
				\cmidrule(lr){3-4}
				& & Calculated & Theoretical & & \\
				\midrule
				\multirow{4}{*}{\makecell{Harmonic \\ Oscillator \\(a.u.)}} 
				& Ground & 0.999994040 & 1.0 & 5.9605$\times 10^{-6}$ & 1.6659$\times 10^{-6}$ \\
				& 1st excited & 1.999990582 & 2.0 & 4.7088$\times 10^{-6}$ & 4.0095$\times 10^{-5}$ \\
				& 2nd excited & 3.000917196 & 3.0 & 3.0573$\times 10^{-4}$ & 9.3415$\times 10^{-4}$ \\
				& 3rd excited & 4.000737190 & 4.0 & 1.8430$\times 10^{-4}$ & 2.1204$\times 10^{-4}$ \\
				\midrule
				\multirow{4}{*}{\makecell{Hydrogen \\ Atom \\(a.u.)}}
				& Ground & -0.498872727 & -1/2 & 2.2545$\times 10^{-3}$ & 1.5221$\times 10^{-4}$ \\
				& 1st excited & -0.124730058 & -1/8 & 2.1595$\times 10^{-3}$ & 6.7828$\times 10^{-5}$ \\
				& 2nd excited & -0.055402730 & -1/18 & 2.7509$\times 10^{-3}$ & 2.4362$\times 10^{-5}$ \\
				& 3rd excited & -0.031143170 & -1/32 & 3.4186$\times 10^{-3}$ & 8.4410$\times 10^{-4}$ \\
				\midrule
				\multirow{4}{*}{\makecell{Charmonium \\ Hadrons \\ (GeV)}}
				& $\eta_c(1^1S_0)$ & 2.983102322 & 2.981 & 7.0524$\times 10^{-4}$ & 6.0181$\times 10^{-4}$ \\
				& $J/\psi(1^3S_1)$ & 3.090467453 & 3.096916 & 2.0822$\times 10^{-3}$ & 2.1677$\times 10^{-4}$ \\
				& $h_c(1^1P_1)$ & 3.516597509 & 3.52541 & 2.4997$\times 10^{-3}$ & 6.1491$\times 10^{-4}$ \\
				& $\chi_c(1P)$ & 3.525164604 & 3.5176 & 2.1505$\times 10^{-3}$ & 2.3918$\times 10^{-4}$ \\
				\midrule
				\multicolumn{6}{l}{\footnotesize Note: The theoretical values for charmonium hadrons are actually experimental measurements.} \\
				\bottomrule
			\end{tabular}
		\end{table}
		
		\begin{figure}[H]
			\centering
			\includegraphics[width=1\textwidth]{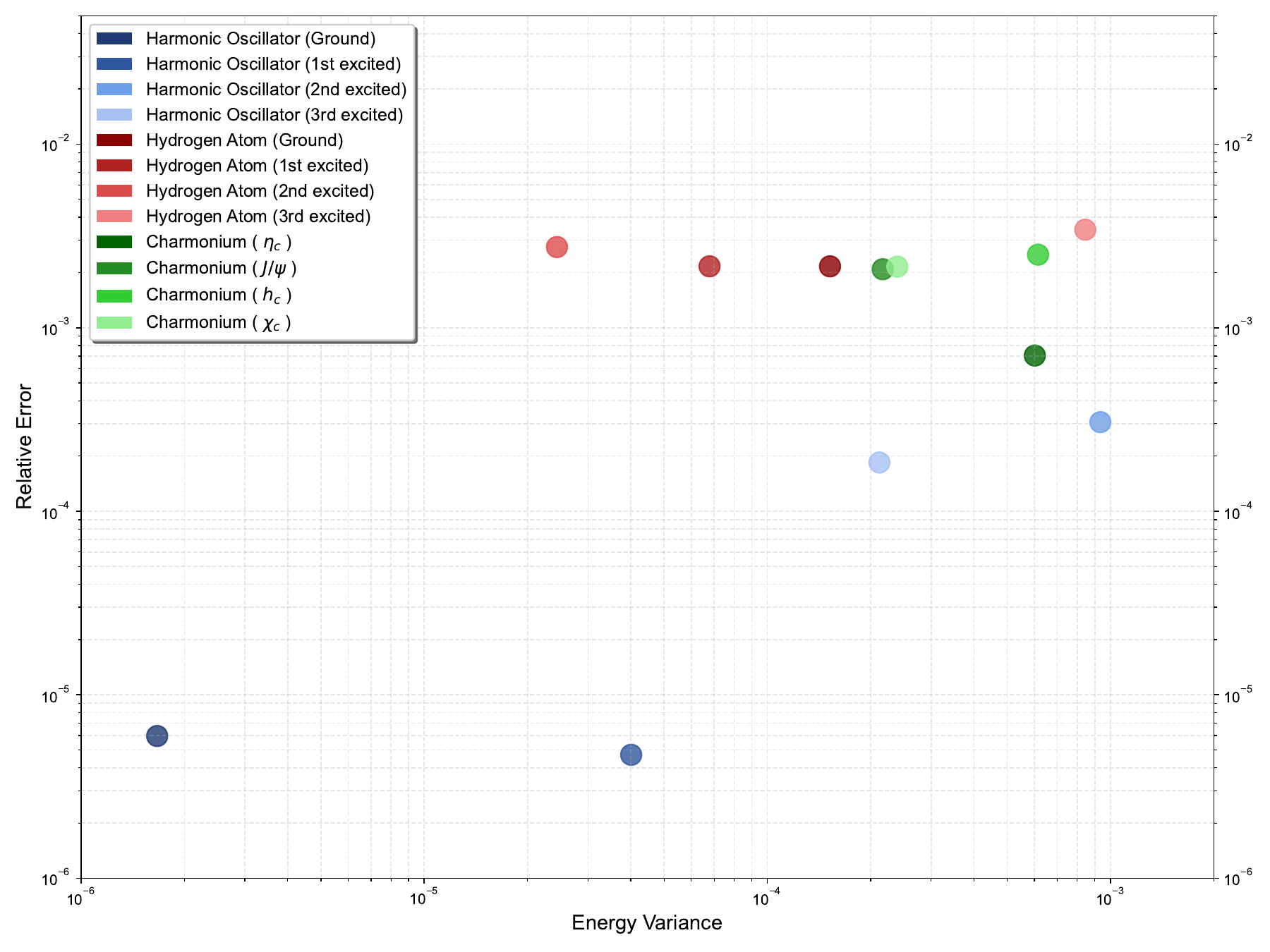}
			\caption{Correlation between energy variance and energy relative error.}
			\label{CorrelationbetweenEA}
		\end{figure}
		
		The relationship between the final energy variance and relative error across all test cases is presented in Fig. \ref{CorrelationbetweenEA}. While a moderate correlation is observed on a logarithmic scale (Pearson correlation coefficient = 0.598), the most significant finding emerges from the empirical distribution of the data points: all systems achieving an energy variance below \(1 \times 10^{-3}\) exhibit relative errors under 1\%. This consistency across systems—from simple harmonic potentials to complex hadronic interactions—demonstrates the broad applicability of our variance-based criterion. We therefore use \(1 \times 10^{-3}\) as a threshold of energy variance for indicating preliminary convergence across diverse quantum systems.This criterion is inherently scalable: demanding higher precision simply requires lowering the variance threshold accordingly.
		
		\begin{figure}[H]
			\centering
			\includegraphics[width=0.9\textwidth]{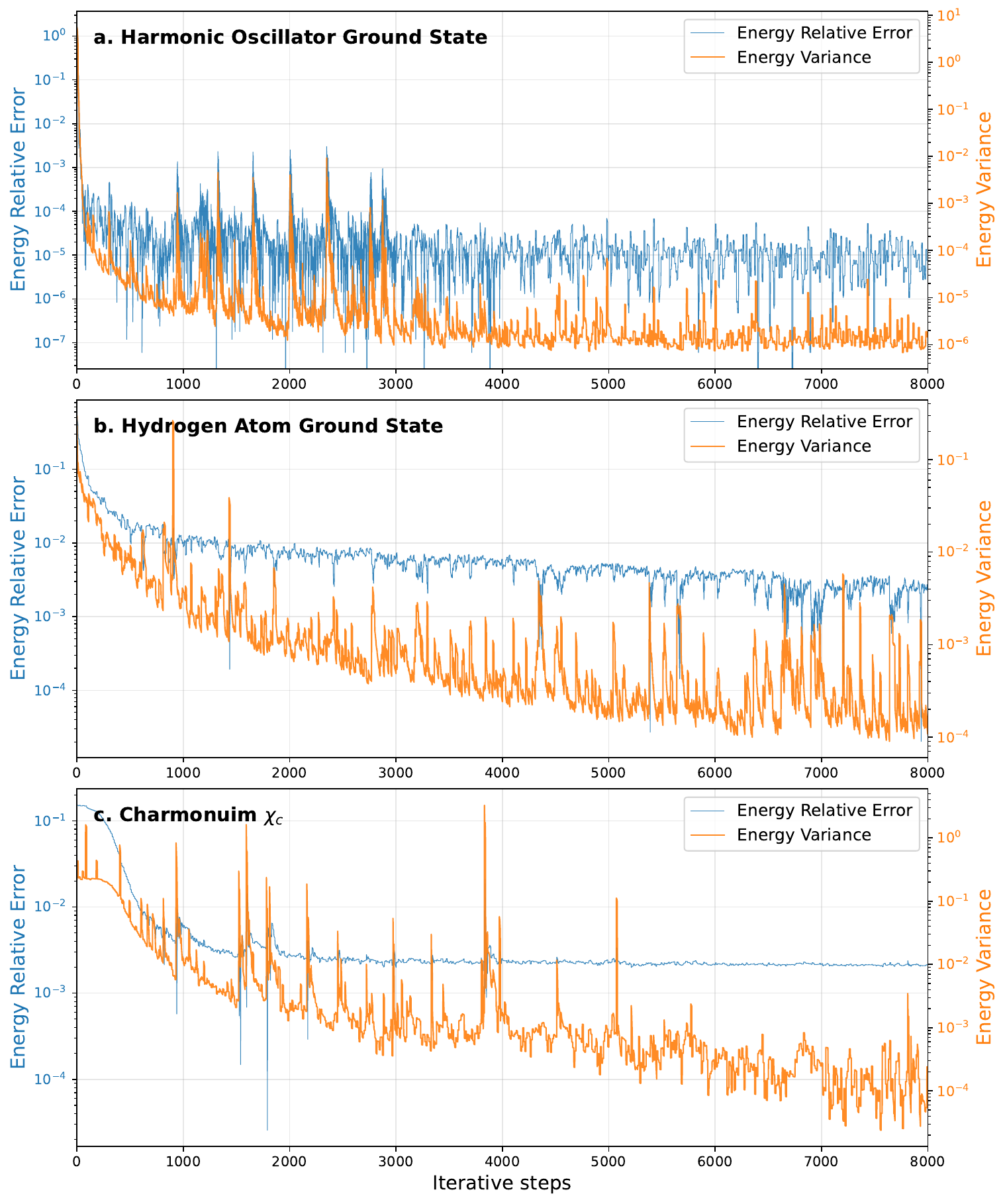}
			\caption{Comparison of energy variance and relative error trajectories.}
			\label{ConvergeImage}
		\end{figure}
		
		The dynamic optimization trajectories of three representative systems—the harmonic oscillator ground state, hydrogen atom ground state, and the $\chi_c$ particle—are shown in Fig. \ref{ConvergeImage}. A key observation is the highly synchronous descent of the energy variance with the energy relative error throughout the training process. This synchronicity confirms that the energy variance serves as a reliable, real-time proxy for the true accuracy of the wave function, even for systems where the exact energy is unknown.
		
		The distinct advantage of the energy variance criterion is further highlighted in the case of the $\chi_c$ particle. As illustrated in Fig. \ref{ConvergeImage}c, the energy curve reaches an apparent plateau, suggesting convergence by a traditional standard. However, the energy variance continues its steady decline, guiding the optimization to a solution of significantly higher accuracy. This demonstrates the criterion's unique ability to identify and overcome deceptive pseudo-convergence, ensuring that the final result represents the best possible solution within the representational capacity of the neural network.
		
		As further corroboration of the physical correctness of the converged results, the wave functions for all cases exhibit the shapes, provided in the \ref{app:figure}.
		
		\subsection{Parameter scanning of multiple systems}
		
		The initial stage of physical research often prioritizes understanding how system properties evolve with external conditions or internal parameters, over pursuing the ultimate precision for a single configuration. A reliable, hands-off convergence criterion is the cornerstone of such large-scale, automated computational campaigns. Leveraging the energy variance criterion, we have developed a lightweight and general-purpose solver whose rapid convergence and low computational footprint make it ideally suited for batch parameter scanning. This capability allows us to efficiently map the ``phase diagrams'' of ground-state properties, providing an intuitive understanding of parametric dependencies and crucial guidance for subsequent focused studies.
		
		\subsubsection{Quantum tunneling in a 2D double-well potential} \label{DoubleWell}
		
		In a symmetric double-well potential separated by a barrier, a classical particle would be localized in one of the wells. However, quantum particles can tunnel through the barrier due to their wave-like nature. This tunneling leads to the formation of delocalized stationary states, with the symmetric state—where the probability density is distributed across both wells—representing the ground state. The Hamiltonian in atomic units is given by:
		
		\begin{equation}
			\hat{H} = -\frac{1}{2m_{\text{eff}}}\nabla^2 + \alpha \left[ x^2 - \left( \frac{d}{2} \right)^2 \right]^2 + \frac{1}{2}m_{\text{eff}}\omega_y^2 y^2,
		\end{equation}
		we set $\alpha = m_{\text{eff}} = 1.0$ and $\omega_y = 2.0$, with the inter-well distance $d$ as the scanning parameter to systematically investigate the tunneling-dependent ground state properties. The calculation data of 2D double-well potential are shown in Table \ref{tab:double_well_scan}.
		
		\begin{table}[htbp]
			\centering
			\caption{Ground state energy of the 2D double-well potential. The relationship between energy and the distance of Well Separation $d$. The consistently low energy variance confirm the reliability of our convergence criterion.}
			\label{tab:double_well_scan}
			\begin{tabular}{ccc}
				\toprule
				Well Separation $d$ (a.u.) & Ground State Energy (a.u.) & Energy Variance \\
				\midrule
				0.0 & 1.167978048 & $9.9352 \times 10^{-7}$ \\
				0.5 & 1.135426760 & $9.0501 \times 10^{-6}$ \\
				1.0 & 1.077275634 & $1.1062 \times 10^{-5}$ \\
				1.5 & 1.105534196 & $8.6934 \times 10^{-5}$ \\
				2.0 & 1.369598150 & $6.9968 \times 10^{-6}$ \\
				2.5 & 1.917830706 & $5.5172 \times 10^{-5}$ \\
				3.0 & 2.470638752 & $8.4875 \times 10^{-6}$ \\
				3.5 & 2.884818316 & $1.7123 \times 10^{-4}$ \\
				4.0 & 3.262554168 & $1.6013 \times 10^{-4}$ \\
				\bottomrule
			\end{tabular}
		\end{table}
		
		\begin{figure}[H]
			\centering
			\includegraphics[width=1\textwidth]{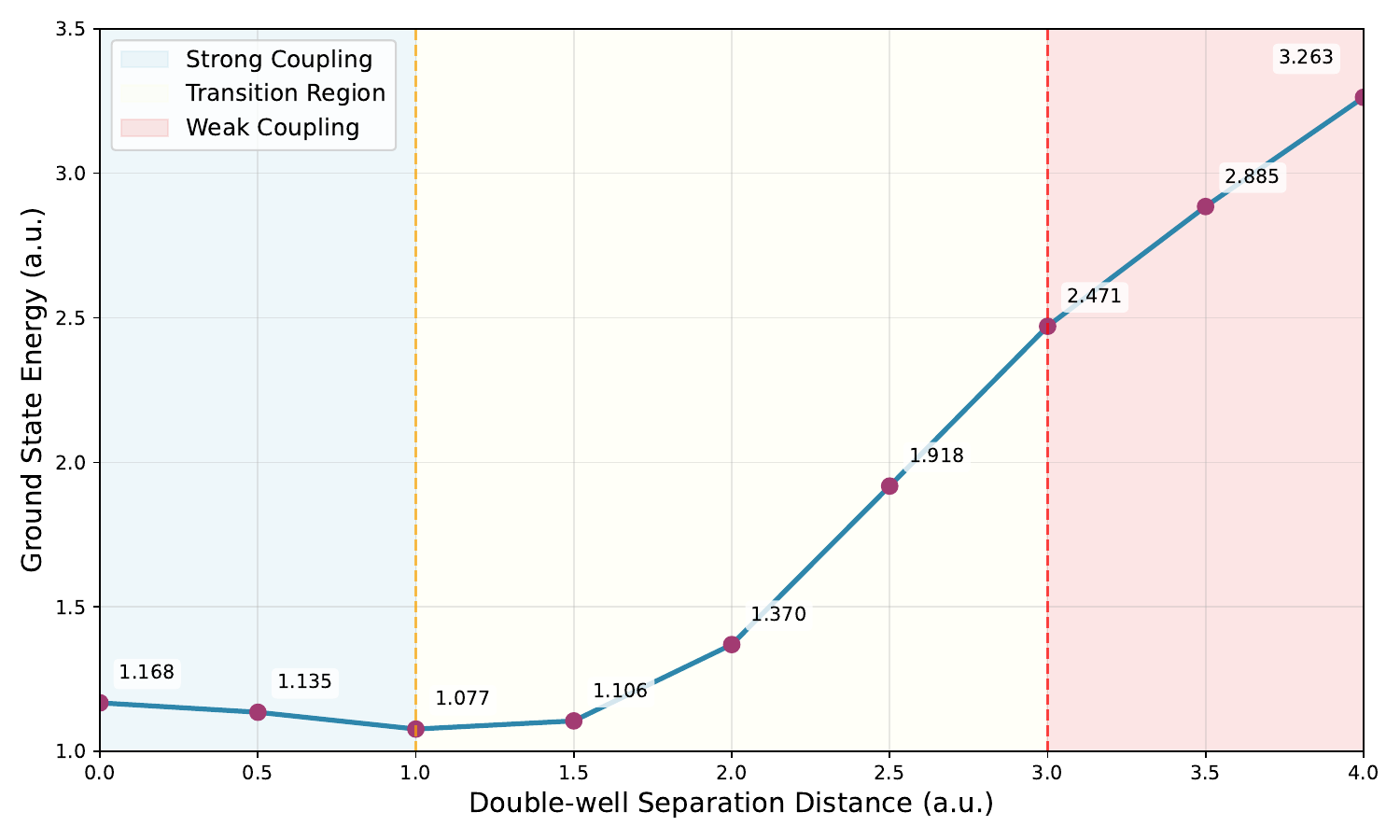}
			\caption{Ground state energy with different separation distance in 2D double-well potential.}
			\label{DoubleWellEnergy}
		\end{figure}
		
		\begin{figure}[H] 
			\centering
			\includegraphics[width=0.8\textwidth]{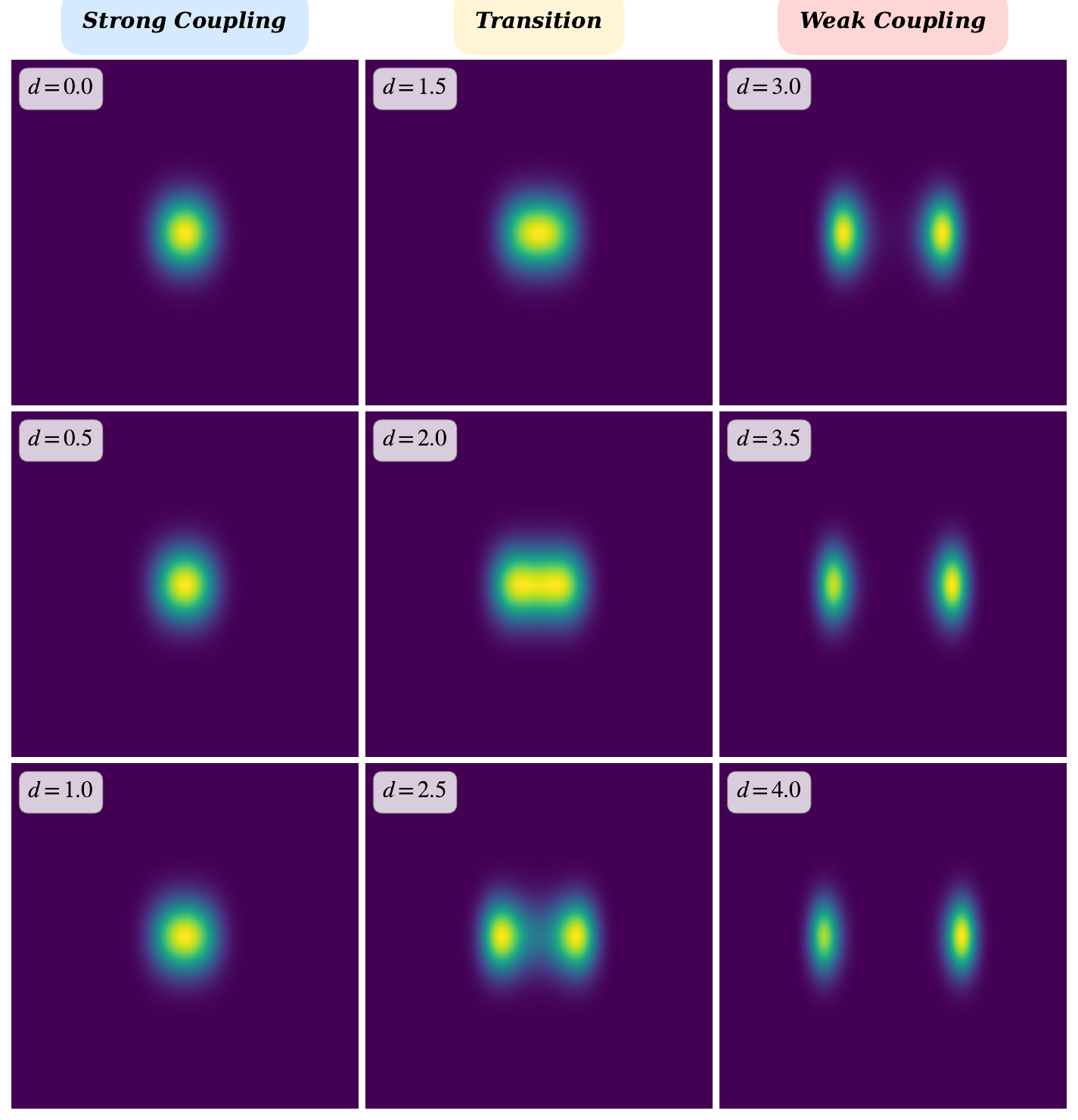}
			\caption{Probability density evolution in 2D double-well potential. The color represents the probability density. The higher the probability density, the brighter the color. The color scale only indicates relative values within the same figure, and the numerical values differ across different figures.}
			\label{DoubleWellWave}
		\end{figure}
		
		The calculated ground states and energy as a function of the well separation \( d \) reveals three distinct regimes, as illustrated in Fig. \ref{DoubleWellEnergy} and Fig. \ref{DoubleWellWave}. Strong coupling regime: An initial increase in the well separation leads to a broadening of the central potential, widening of the wavefunction, and a reduction in the kinetic energy contribution, thereby lowering the ground state energy. In this regime, the wavefunction exhibits an undivided, single-peaked or nearly single-peaked morphology. Transition regime: A turning point is reached, beyond which the energy increases monotonically with \(d\). The wavefunction begins to separate significantly, forming a stable double-peaked structure, accompanied by a decline in the overlap and coupling between the left and right localized states. Weak coupling regime: The energy curve maintains a positive slope, but the growth rate diminishes, reflecting the system's approach towards a configuration of two independent harmonic oscillators. The two wells become nearly independent with extremely weak coupling.
		
		\subsubsection{Compression of the hydrogen atom in the strong magnetic field}
		
		The ground state of the hydrogen atom is robust against weak magnetic fields. To manifest the magnetic field's influence distinctly, we apply strong magnetic fields along the z-axis, with strengths ranging from magnetic flux density \(B = 0.1\) to \(1.0\) atomic units (where 1 a.u. \(\approx 2.35 \times 10^5\) T, far exceeding the highest magnetic field on Earth). This allows us to clearly observe the impact on both the ground-state energy and the wave function morphology. 
		
		For the ground state, whose magnetic quantum number is zero, \(L_z\) is zero and therefore the paramagnetic term vanishes, leaving only the diamagnetic term in the Hamiltonian. The Hamiltonian in atomic units is given by:
		
		\begin{equation}
			\hat{H} = -\frac{1}{2}\nabla^2 - \frac{1}{r} + \frac{B^2}{8}(x^2 + y^2),
		\end{equation}
		where $r = \sqrt{x^2+y^2+z^2}$ . The calculation data are shown in Table \ref{tab:magnetic_scan}.
		
		\begin{table}[htbp]
			\centering
			\caption{Ground state energy of the hydrogen atom under varying magnetic field strengths.}
			\label{tab:magnetic_scan}
			\begin{tabular}{ccc}
				\toprule
				Magnetic Flux Density \(B\) (a.u.) & Ground State Energy (a.u.) & Energy Variance \\
				\midrule
				0.1 & -0.495588571 & $8.1131 \times 10^{-4}$ \\
				0.2 & -0.488055795 & $3.2198 \times 10^{-4}$ \\
				0.3 & -0.477083296 & $2.7810 \times 10^{-4}$ \\
				0.4 & -0.461531669 & $3.5817 \times 10^{-4}$ \\
				0.5 & -0.443386376 & $4.2902 \times 10^{-4}$ \\
				0.6 & -0.423752874 & $5.1532 \times 10^{-4}$ \\
				0.7 & -0.403064519 & $3.0886 \times 10^{-4}$ \\
				0.8 & -0.379649967 & $1.5625 \times 10^{-4}$ \\
				0.9 & -0.352227509 & $8.1564 \times 10^{-4}$ \\
				1.0 & -0.327324480 & $9.0777 \times 10^{-4}$ \\
				\bottomrule
			\end{tabular}
		\end{table}
		
		\begin{figure}[H] 
			\centering
			\includegraphics[width=1\textwidth]{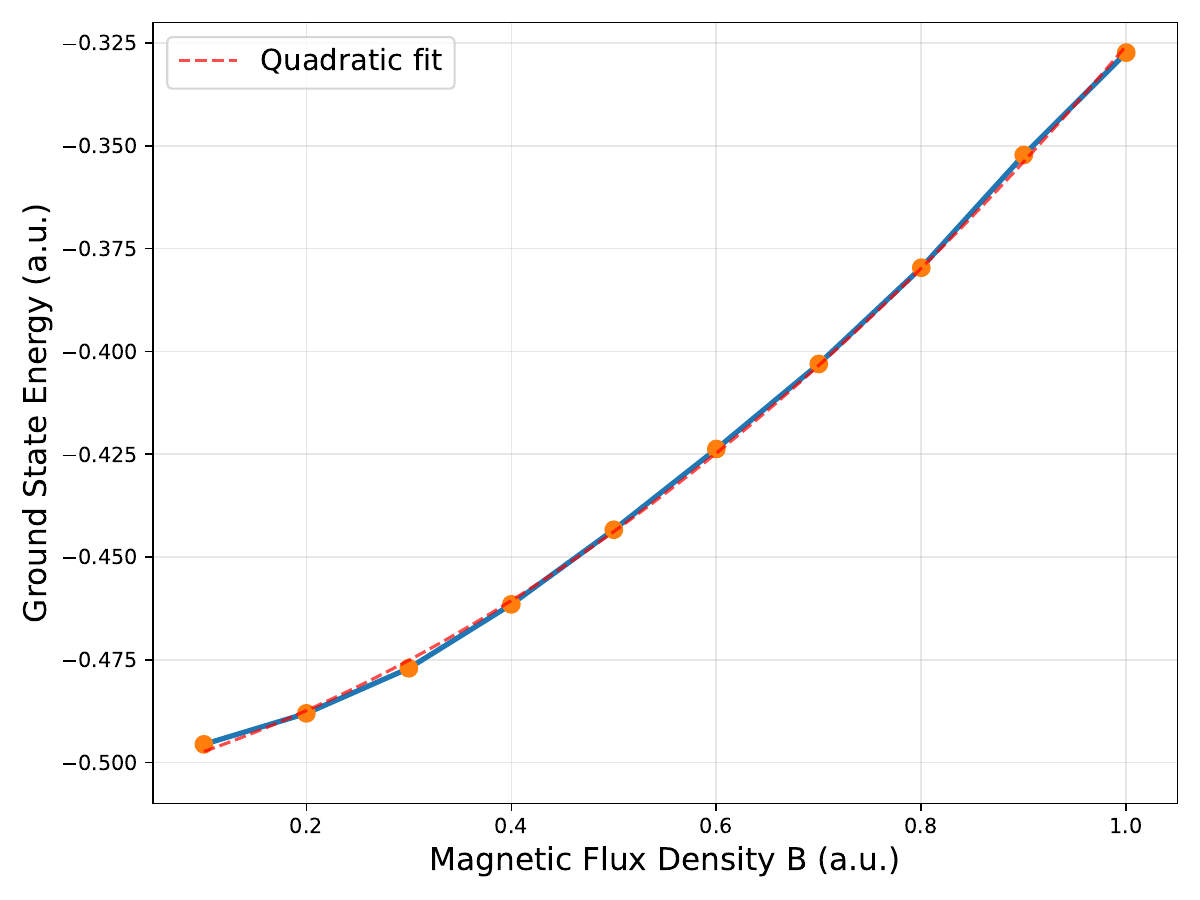}
			\caption{Hydrogen atom ground state energy variation with magnetic field strength.}
			\label{StrongMagfield}
		\end{figure}
		
		Plot the curve of the system energy versus magnetic induction intensity and fit it with a quadratic polynomial as shown in Fig. \ref{StrongMagfield}. The ground state energy increases monotonically with the magnetic field strength, rising from -0.4956 a.u. at \(B = 0.1\) a.u. to -0.327 a.u. at \(B = 1.0\) a.u.. This trend aligns well with theoretical predictions and is primarily attributed to the positive energy contribution from the diamagnetic term in the Hamiltonian.
		
		\begin{figure}[H] 
			\centering
			\includegraphics[width=1\textwidth]{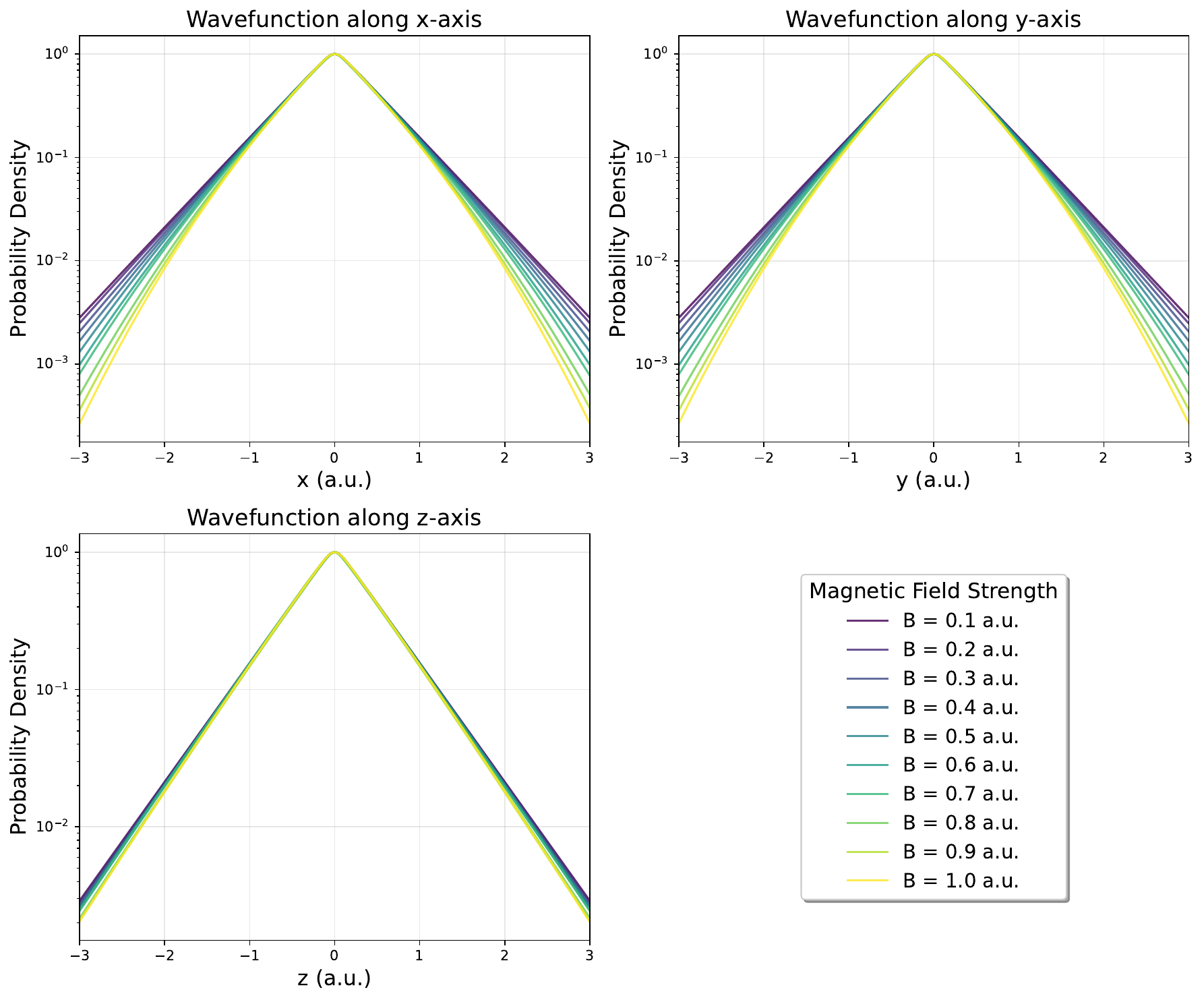}
			\caption{Changes in hydrogen atom ground state wavefunction under different magnetic field strengths. }
			\label{MagneticFields3D}
		\end{figure}
		
		The changes in the wave function are relatively subtle shown as Fig. \ref{MagneticFields3D}. By employing a logarithmic scale for visualization, it becomes apparent that the electron density exhibits significant compression in the xy-plane (perpendicular to the magnetic field) as the field strength increases. Furthermore, we observe a slight, systematic contraction along the z-axis. This indicates that under a strong magnetic field, the electron cloud of the hydrogen atom undergoes a comprehensive three-dimensional redistribution, rather than being affected only in the plane perpendicular to the field.
		
		\subsubsection{Three-body quantum dot in a anisotropic harmonic oscillator trap}
		
		This section investigates the ground state of a three-body quantum dot with interacting particles by scanning the harmonic confinement strengths along different spatial directions. The Hamiltonian in atomic units is given by:
		
		\begin{equation}
			\hat{H} = \sum_{i=1}^3 \left[ -\frac{1}{2m_{\text{eff},i}} \nabla_i^2 + \frac{1}{2} m_{\text{eff},i} \left( \omega_x^2 x_i^2 + \omega_y^2 y_i^2 + \omega_z^2 z_i^2 \right) \right] + \sum_{j<k} \frac{1}{r_{jk}},
		\end{equation}
		where $r_{jk}=\sqrt{(x_j-x_k)^2+(y_j-y_k)^2+(z_j-z_k)^2}$, and the effective mass is set to \( m_{\text{eff}} = 1.0 \) for all particles. The confinement frequencies \( \omega_x \) and \( \omega_y \) are treated as scanning parameters to analyze their influence on the system's ground state. The calculation data are shown in Table \ref{tab:quantum_dot_energy} and \ref{tab:quantum_dot_variance}.
		
		\begin{table}[htbp]
			\centering
			\caption{Ground-state energies of the three-body quantum dot under varying confinement strengths $\omega_x$ and $\omega_y$. All values are given in atomic units.}
			\label{tab:quantum_dot_energy}
			\begin{tabular}{c|ccccc}
				\toprule
				\multirow{2}{*}{$\omega_y$} & \multicolumn{5}{c}{$\omega_x$} \\
				\cmidrule(lr){2-6}
				& 0.10 & 0.15 & 0.20 & 0.25 & 0.30 \\
				\midrule
				0.10 & 1.016510606 & 1.126879096 & 1.218364358 & 1.302534461 & 1.384103537 \\
				0.15 & 1.126871347 & 1.250050545 & 1.347112894 & 1.435063004 & 1.518608332 \\
				0.20 & 1.217986345 & 1.347136974 & 1.447034240 & 1.537023544 & 1.622270823 \\
				0.25 & 1.302550912 & 1.434744835 & 1.536868930 & 1.627518654 & 1.713410497 \\
				0.30 & 1.384493470 & 1.519017816 & 1.621944904 & 1.713231206 & 1.799155951 \\
				\bottomrule
			\end{tabular}
		\end{table}
		
		\begin{table}[htbp]
			\centering
			\caption{Energy variance of the three-body quantum dot under varying confinement strengths $\omega_x$ and $\omega_y$.}
			\label{tab:quantum_dot_variance}
			\begin{tabular}{c|ccccc}
				\toprule
				\multirow{2}{*}{$\omega_y$} & \multicolumn{5}{c}{$\omega_x$} \\
				\cmidrule(lr){2-6}
				& 0.10 & 0.15 & 0.20 & 0.25 & 0.30 \\
				\midrule
				0.10 & 2.4165$\times 10^{-4}$ & 2.5429$\times 10^{-4}$ & 4.5772$\times 10^{-4}$ & 2.8428$\times 10^{-4}$ & 7.0386$\times 10^{-4}$ \\
				0.15 & 3.5265$\times 10^{-4}$ & 4.1570$\times 10^{-4}$ & 3.4838$\times 10^{-4}$ & 7.1356$\times 10^{-4}$ & 3.6103$\times 10^{-4}$ \\
				0.20 & 3.7541$\times 10^{-4}$ & 5.1044$\times 10^{-4}$ & 3.6602$\times 10^{-4}$ & 7.8924$\times 10^{-4}$ & 4.8291$\times 10^{-4}$ \\
				0.25 & 1.5092$\times 10^{-4}$ & 4.0504$\times 10^{-4}$ & 6.3797$\times 10^{-4}$ & 6.2048$\times 10^{-4}$ & 4.8457$\times 10^{-4}$ \\
				0.30 & 8.9360$\times 10^{-4}$ & 5.3790$\times 10^{-4}$ & 7.0146$\times 10^{-4}$ & 6.6746$\times 10^{-4}$ & 4.0230$\times 10^{-4}$ \\
				\bottomrule
			\end{tabular}
		\end{table}

		\begin{figure}[H] 
			\centering
			\includegraphics[width=1\textwidth]{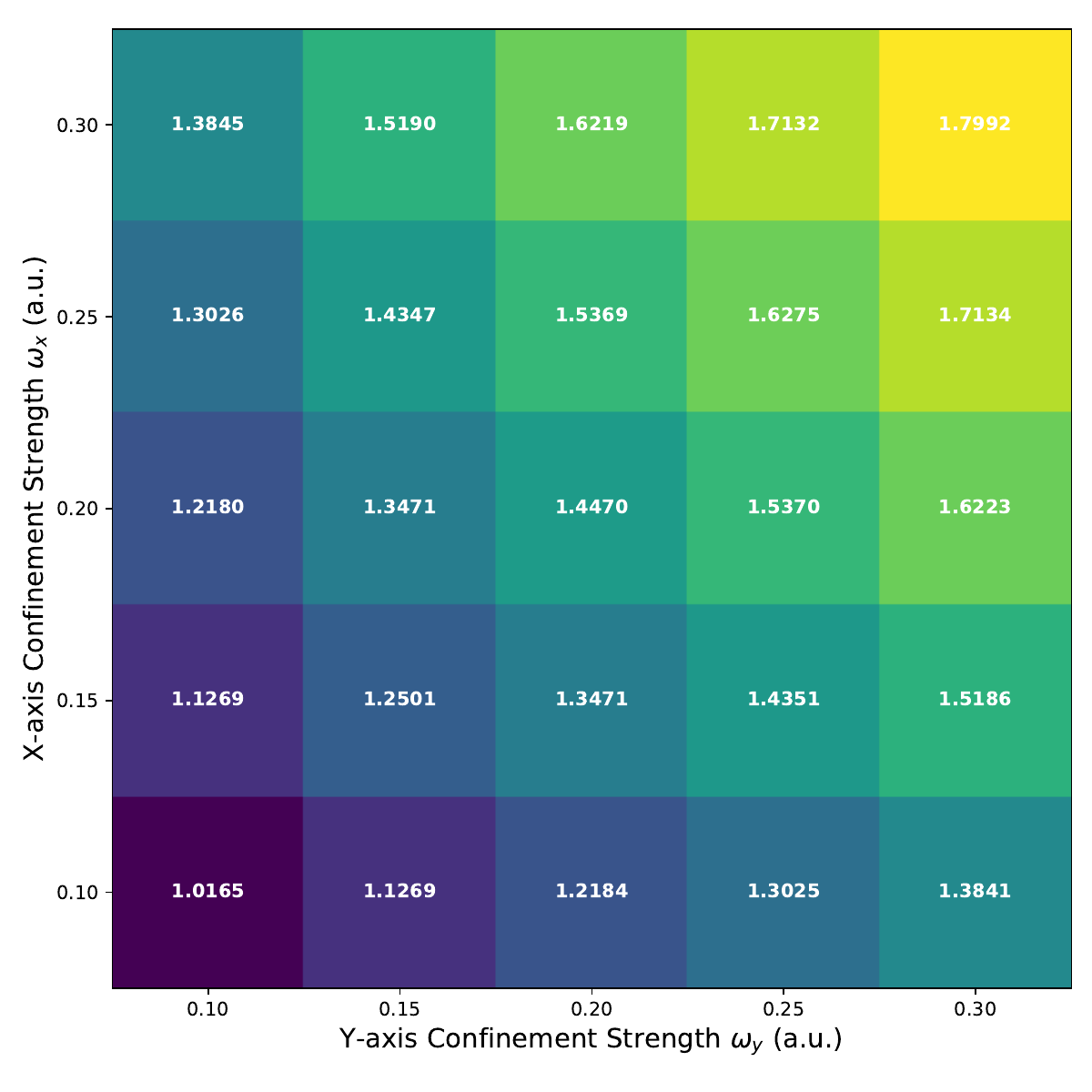}
			\caption{Three-body quantum dot ground state energy in anisotropic harmonic oscillator trap.}
			\label{ThreeBodyHeat}
		\end{figure}
		
		Plot the variation of energy with confinement strength as Fig. \ref{ThreeBodyHeat}. The energy heat map clearly reveals the system's symmetry and monotonicity: energy increases monotonically with confinement strength. The symmetry along the diagonal ($\omega_x=\omega_y$) indicates consistent system response to confinement in both directions, while slight asymmetry in off-diagonal regions reveals anisotropic effects. The energy variance, on the other hand, displays no such pattern. This absence of correlation is a strength of the criterion: it confirms that the variance serves as a robust and unbiased metric purely for convergence, independent of the specific physical changes induced by the parameters. 
		
		\begin{figure}[H] 
			\centering
			\includegraphics[width=1\textwidth]{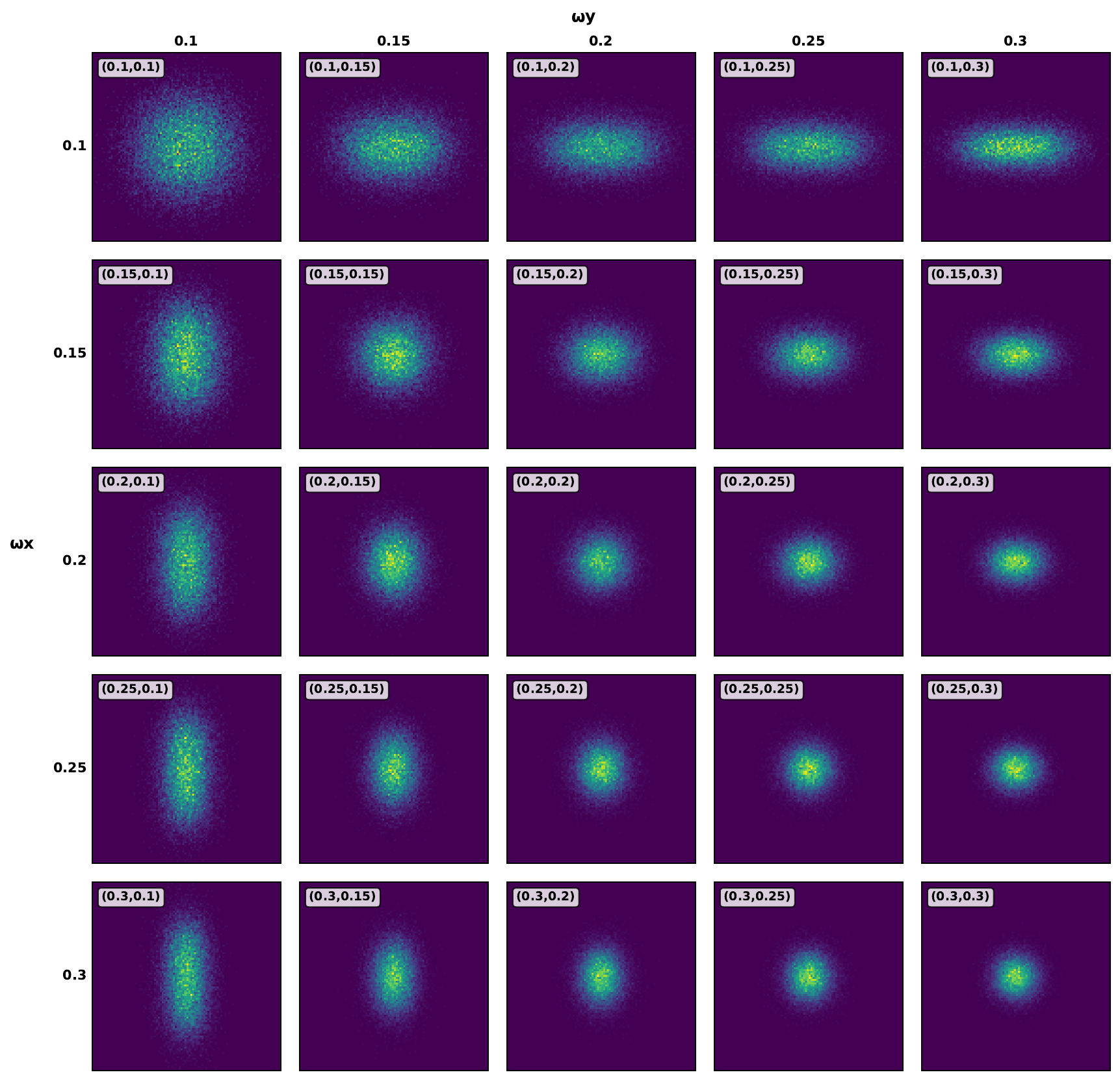}
			\caption{Single particle marginal density distribution in XY plane. The color represents the probability density. The higher the probability density, the brighter the color. The color scale only indicates relative values within the same figure, and the numerical values differ across different figures.}
			\label{QuantumDotGrid}
		\end{figure}
		
		We also show the single particle marginal density as Fig. \ref{QuantumDotGrid}, which illustrate the impact of confinement strength on the quantum dot's spatial structure. In the weak confinement region, the electron cloud is significantly extended. As the confinement strength increases, the electron cloud is progressively compressed, forming distinct ellipsoidal shapes under anisotropic conditions ($e.g.,\omega_x=0.10,\omega_y=0.30$).
		
		\begin{figure}[H] 
			\centering
			\includegraphics[width=0.7\textwidth]{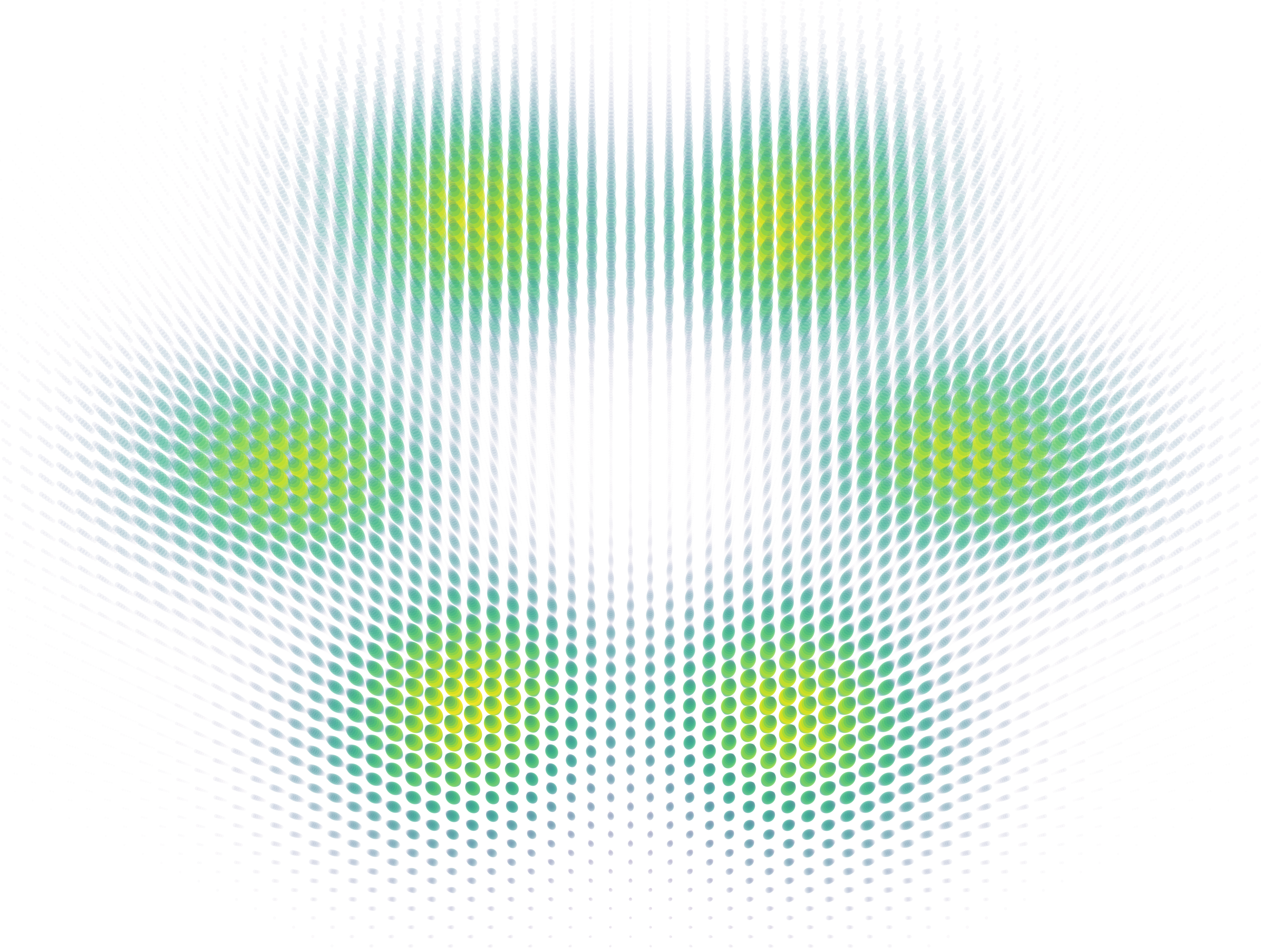}
			\caption{The probability density of interaction space composed of the x-coordinates of three particles.}
			\label{correlation}
		\end{figure}
		
		Plot the probability density in the interaction space using the x-coordinates of the three particles as the coordinate axes, as shown in Fig. \ref{correlation}. The six-fold symmetric wave function structure observed in the interaction space reflects the quantum correlation characteristics of the three-particle system. These six symmetrically distributed ``elliptical lobes'' in the center-of-mass plane correspond to specific modes of relative particle motion. This complex structure originates from repulsive interactions between particles, signaling the system's entry into a strongly correlated quantum state.
		
		\begin{figure}[H] 
			\centering
			\includegraphics[width=1\textwidth]{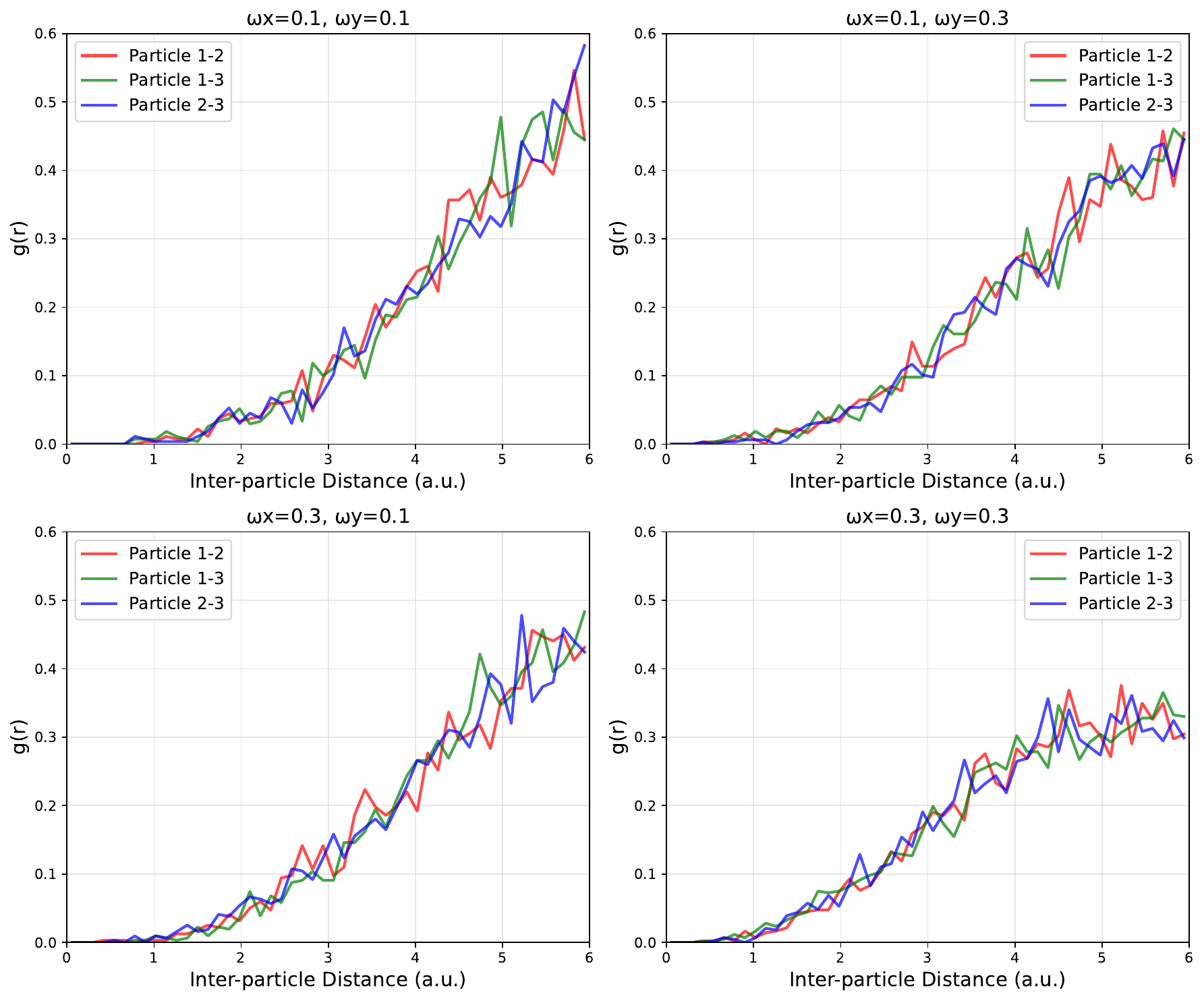}
			\caption{Detailed pair correlation function (three particles pairs).}
			\label{DetailedPairCorrelations}
		\end{figure}
		
		We analyzed the two-body correlation function in the three-body quantum dot shown as Fig. \ref{DetailedPairCorrelations}, examining the distance distributions for three particle pairs under four typical parameter sets. The two body correlation function can be defined as:
		
		\begin{equation}
			g_{ij}(r) = C\left\langle \delta(r - |\mathbf{r}_i - \mathbf{r}_j|) \right\rangle,
		\end{equation}
		where $C$ is normalization factor. It describes the probability density of this particle pair occurring at different inter-particle distances.
		
		The three correlation function curves overlap highly, confirming the system's excellent permutation symmetry. Images from different confinement strength combinations show that anisotropic confinement conditions affect all three particle pairs equally, and higher confinement strength leads to particles tending to be closer together.

		\subsection{Discussion on the robustness and limitations of the energy variance convergence criterion}
		
		While the energy variance criterion is theoretically rigorous, its practical implementation requires careful consideration of numerical and physical complexities. A key concern for any convergence criterion is the occurrence of false positives—where the criterion is met, but the solution is physically incorrect. In our experiments, a salient example was observed in the weakly-coupled double-well potential at Sec. \ref{DoubleWell}.
		
		\begin{figure}[H] 
			\centering
			\includegraphics[width=1\textwidth]{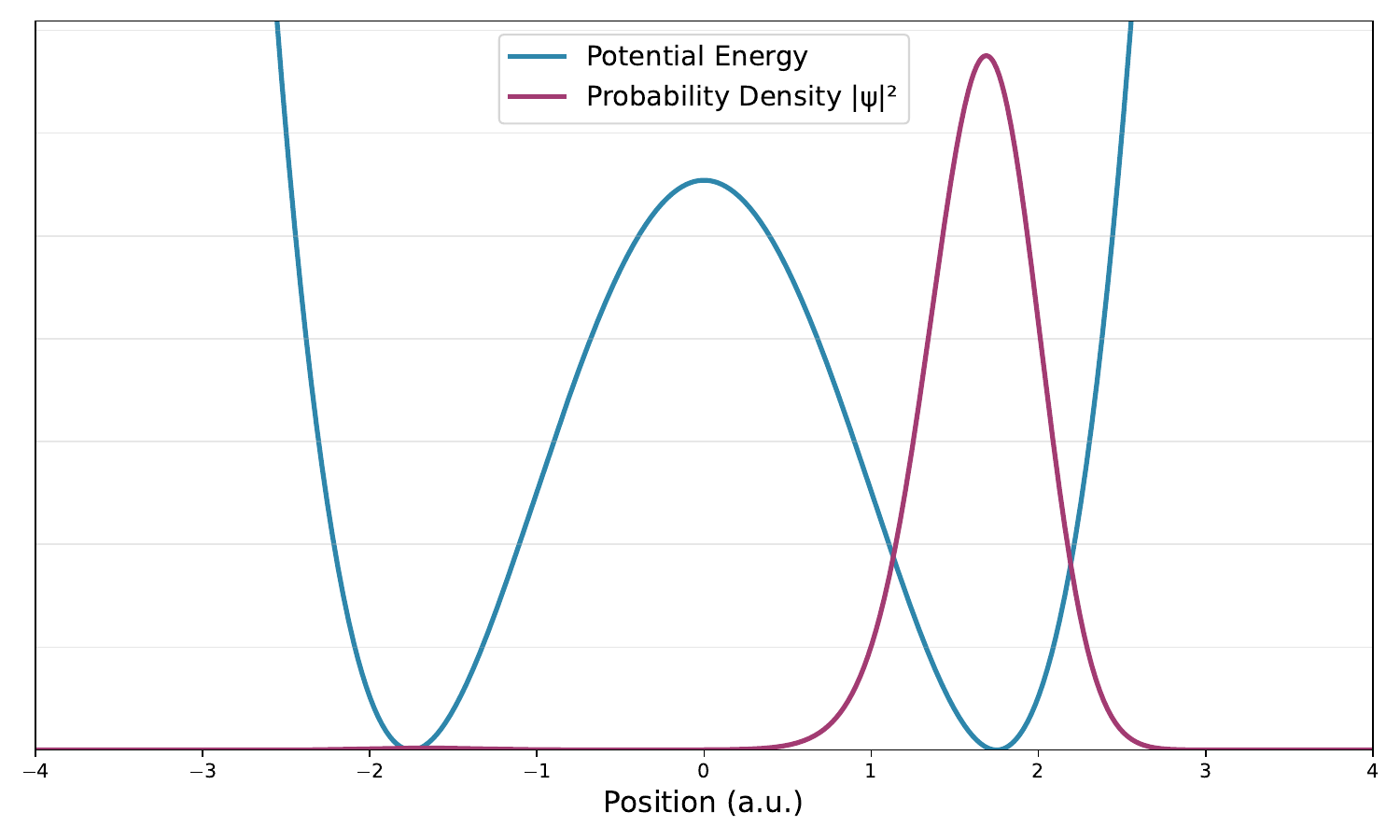}
			\caption{Localized state in 2D double-well potential.}
			\label{LocalizedWave}
		\end{figure}
		
		A representative case of convergence to a localized state at \(d = 3.5\) a.u. is presented in Fig. \ref{LocalizedWave}. Despite its low energy variance (\(4.073 \times 10^{-5}\)), this state is not the true symmetric ground state. This occurs because, at large separations, the system exhibits near-degeneracy: the energy splitting between the symmetric and symmetry-broken states becomes exceedingly small, comparable to or smaller than the numerical precision of the training. Therefore, the neural network cannot effectively resolve the true ground state and may stabilize in a localized configuration.
		
		This instance highlights a fundamental trade-off between different convergence philosophies. While powerful frameworks like NetKet\cite{vicentiniNetKet3Machine2022} employ sophisticated statistical diagnostics like the Gelman-Rubin statistic to ensure robust sampling—enabling them to tackle extremely challenging systems—our approach prioritizes physical interpretability and computational agility. The energy variance provides a direct, quantum-mechanical measure of wave function quality, and its implementation requires only a single optimization run. This makes our paradigm particularly suited for the rapid preliminary investigation of parameterized Hamiltonians. The two approaches are complementary: our variance criterion could, in future work, be integrated as a high-level physical validation step within more complex frameworks that use statistical diagnostics for low-level sampling control.

		\section{Conclusion} \label{Sec4}
		
		In this work, we have established the energy variance as a robust and practical convergence criterion for neural wave function optimization in variational Monte Carlo. By translating the fundamental quantum-mechanical principle—that eigenstates exhibit zero variance—into an automated computational framework, we have developed a lightweight neural solver that enables efficient and reliable calculations for quantum systems. A theoretical analysis of the variance landscape further clarifies the domain of applicability of our criterion, while comprehensive numerical validations confirm its effectiveness across diverse systems. Our approach consistently delivers accurate results with modest computational resources, underscoring its strengths in accessibility and scalability.
		
		Across a range of non-nodal ground-state systems—including harmonic oscillators, hydrogen atoms, and charmonium—variance minimization proves both effective and reliable. We further propose an empirical variance threshold of \(10^{-3}\) as a convergence standard, which facilitates automated parameter scans and significantly accelerates the preliminary physical verification process. It should be noted that the precise value of this threshold may vary with system size and complexity. The successful application of the variance as a convergence diagnostic in excited states further demonstrates that, although variance minimization may not always serve as a stable optimization parameter in nodal structures, the variance itself remains a valuable indicator of convergence.
		
		Looking forward, future work will focus on extending the applicability of our framework by developing technical strategies to mitigate singularities, thereby effectively generalizing the energy-variance criterion to nodal systems. This advancement will further broaden the utility of our approach to strongly correlated and complex fermionic systems.

		\section*{Acknowledgements}
		
		We thank Ling-Da Ruan for useful discussions. This work is supported by the National Natural Science Foundation of China under Grants No.~12005172 and the Fundamental Research Funds for the Central Universities. 
		
		\section*{Data Availability Statement}
		
		The data that support the findings of this study are available within the article and its Supplementary Material. The source code used in this study will be made publicly available on GitHub upon publication.
		
		\bibliographystyle{unsrt}
		\bibliography{references}
		
		\appendix
		\section{Wave function visualizations}
		\label{app:figure}
		
		This appendix provides visualizations of the wave functions for the benchmark quantum systems discussed in the Sec. \ref{Validation}, serving as further validation of the physical correctness of our computational results. All wave functions were optimized by our solver under the guidance of the energy-variance convergence criterion. Their morphologies are in agreement with the theoretical expectations for their respective systems. 
		
		\begin{figure}[H] 
			\centering
			\includegraphics[width=0.5\textwidth]{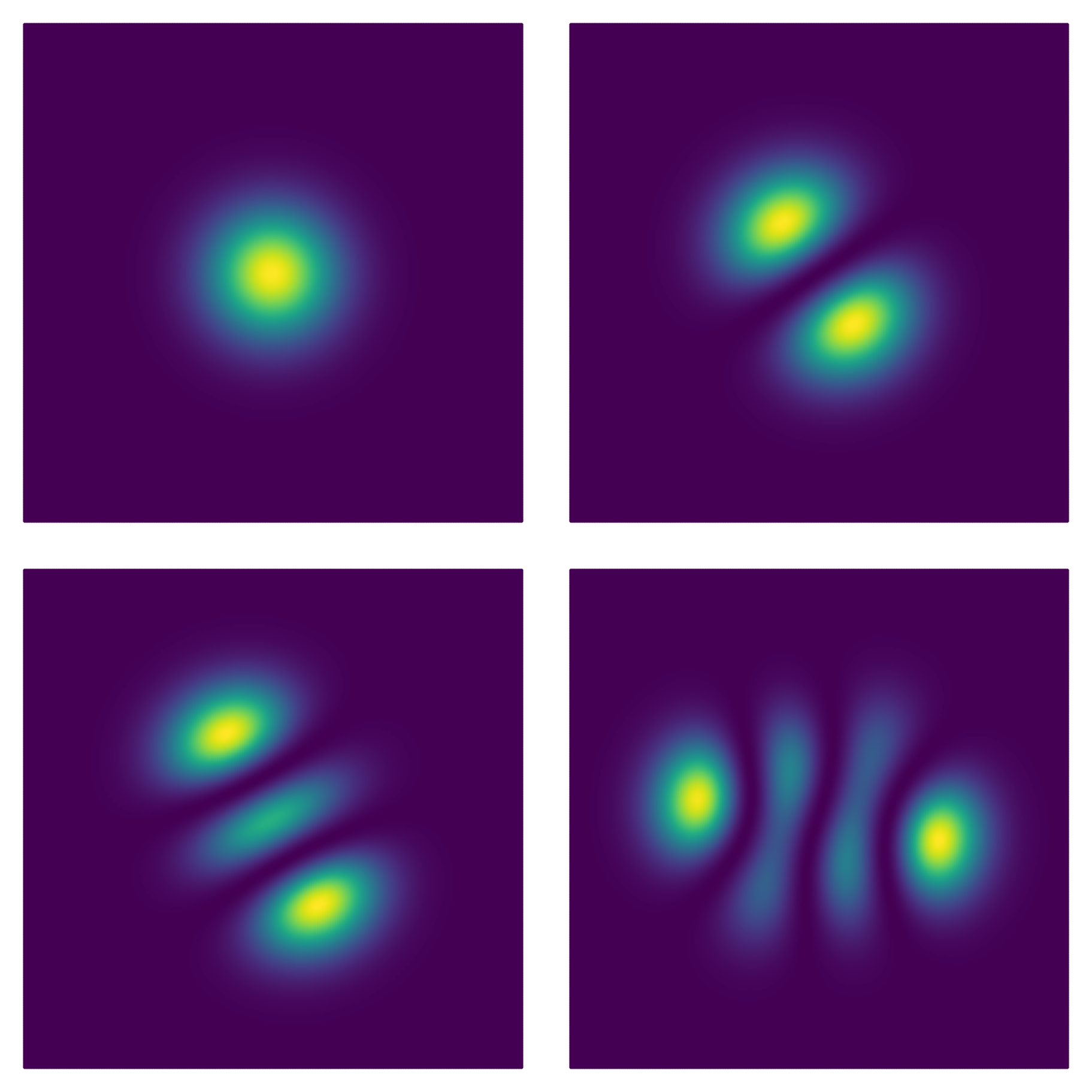}
			\caption{Probability density of the harmonic oscillator.}
			\label{HarmonicOscillatorWave}
		\end{figure}
		
		The probability density for the first four energy eigenstates (ground, first, second, and third excited states) of the two-dimensional harmonic oscillator shown as Fig. \ref{HarmonicOscillatorWave}. The increasing number of nodes with energy is clearly observed. The symmetry and spatial distribution of these states match the exact theoretical solutions perfectly, demonstrating the solver's accuracy in capturing wave functions within a harmonic potential.
		
		\begin{figure}[H] 
			\centering
			\includegraphics[width=0.5\textwidth]{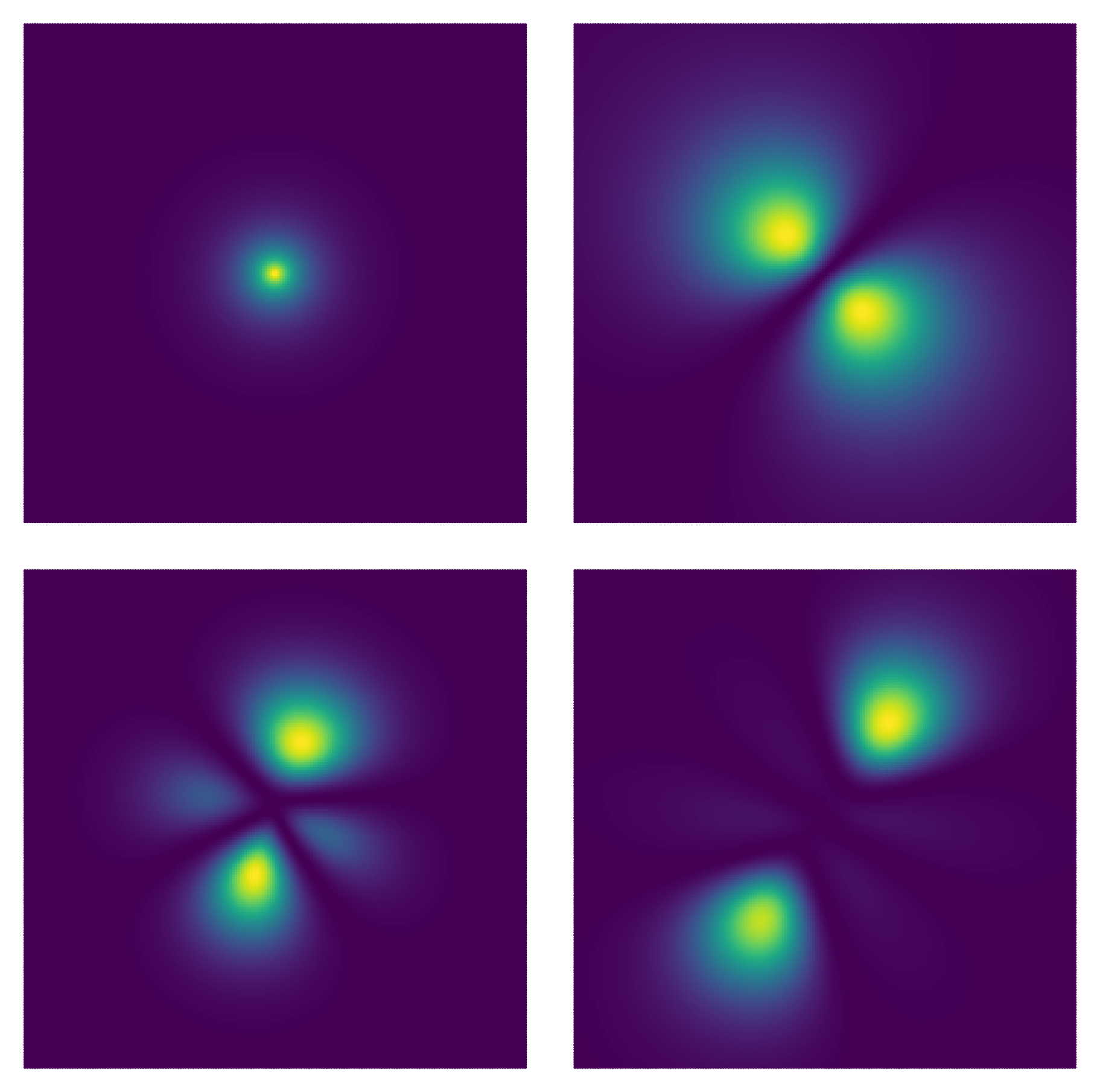}
			\caption{Probability density of the Hydrogen atom.}
			\label{HydrogenAtomWave}
		\end{figure}
		
		The computed low-lying states of the hydrogen atom shown as Fig. \ref{HydrogenAtomWave}. Due to the inherent degeneracy within each principal quantum number \( n \) in the Coulomb potential, the neural network converges to specific angular momentum states. Notably, the optimizer consistently selects states with high orbital angular momentum, yielding the \( 1s \), \( 2p \), \( 3d \), and \( 4f \) orbitals. The faint lobes in the computed \( 4f \) orbital are attributable to the current numerical precision, which is sufficient for accurate energy estimation but leaves room for improvement in fully resolving the most complex details of high-\( l \) states.
		
		\begin{figure}[H] 
			\centering
			\includegraphics[width=0.9\textwidth]{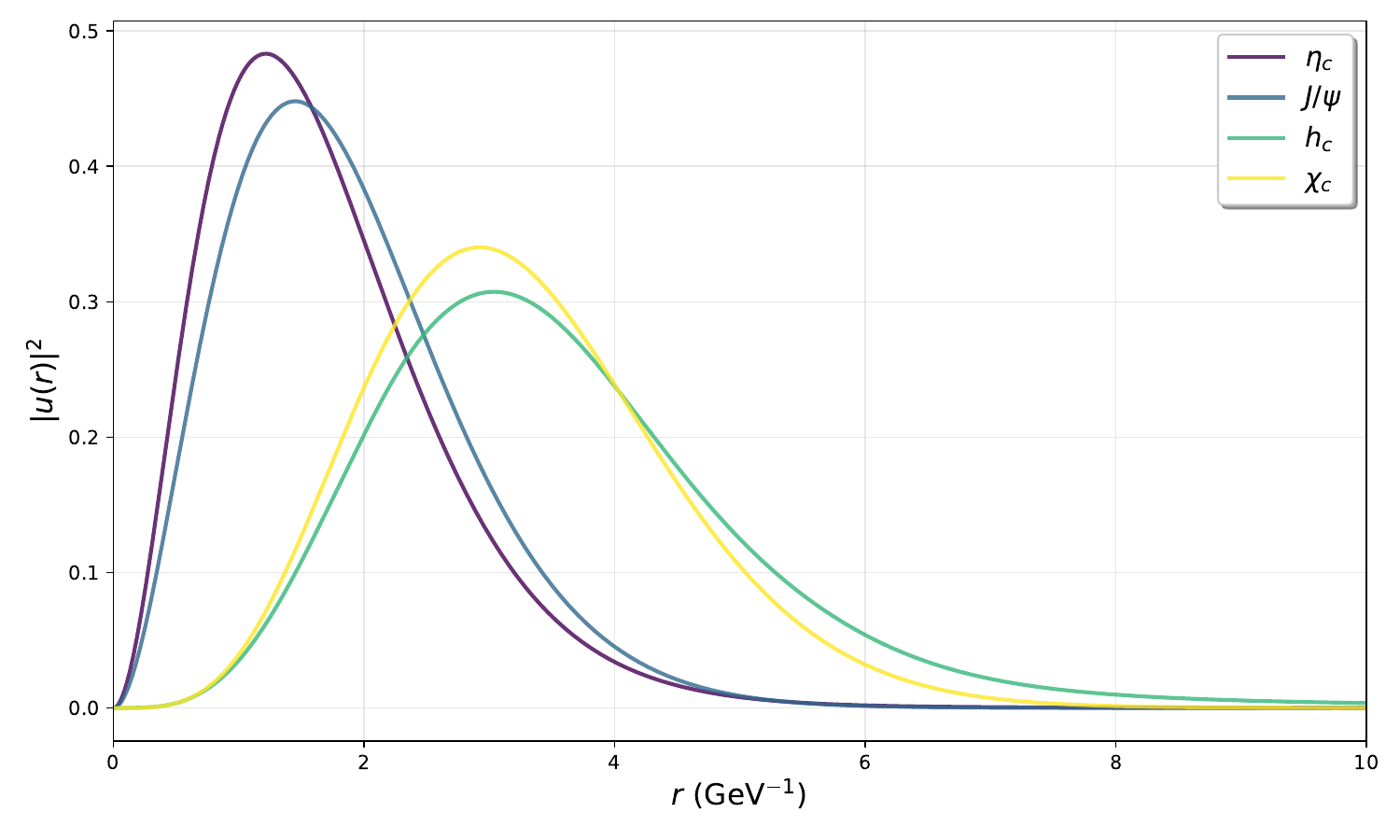}
			\caption{Reduced radial probability density of Charmonium states.}
			\label{MeasonWave}
		\end{figure}
		
		The reduced radial probability density shown as Fig. \ref{MeasonWave}, \( u(r) = r R(r) \), for the four charmonium states (\(\eta_c\), \(J/\psi\), \(h_c\), \(\chi_c\)). Their shapes and spatial extents correctly reflect the physical characteristics under different orbital angular momenta and spin configurations, confirming our method's capability for handling complex interaction potentials.

	\end{document}